\documentclass[12pt]{article}
\begin{document}
%\draft
\hfill\vbox{\baselineskip14pt
            \hbox{\bf November 2004}}
%            \hbox{\bf ETL-00-xxx}
%            \hbox{ETL Preprint 00-xxx}
%            \hbox{\today}}
%            \hbox{January 1998}}
%            \hbox{July 2003}}
\baselineskip20pt
\vskip 0.2cm 
\begin{center}
%{\Large\bf  Electronic Structure of PbVO$_{3}$ studied by XANES }
%\end{center} 
{\Large\bf  Local Electronic Structure of PbVO$_{3}$, a New Member of
PbTiO$_{3}$ Family, studied by XANES/ELNES }
\end{center}
%\vskip 0.2cm 
\begin{center}
\large Sher~Alam$^{1}$, Alexei A. Belik$^{2}$, Y. Matsui$^{1}$
 %%%%%%%%%%%%%
%\footnote{Permanent address: Department of Physics, University
%of Peshawar, Peshawar, NWFP, Pakistan.}
\end{center}
\begin{center}
$^{1}${\it CSAAG, NIMS, Tsukuba 305-0044, Ibaraki, Japan}\\
$^{2}${\it ICYS, NIMS, Tsukuba 305-0044, Ibaraki, Japan}\\
%$^{3}${\it Nanoelectronics, Nat.~Inst.~of~AIST, Tsukuba 305-8568,
%Ibaraki, Japan}\\
%$^{4}${\it Kandhar Virtual University, 448-18 Ozone, Tsukuba, Ibaraki, Japan}
\end{center}
%\begin{center}
%{\it Physical Science Division, ETL, Tsukuba, Ibaraki 305, Japan}
%\end{center}
%\vskip 0.2cm 
\begin{center} 
\large Abstract
\end{center}
\begin{center}
%\begin{minipage}{14cm}
%\begin{minipage}{18cm}
\begin{minipage}{16cm}
\baselineskip=18pt
\noindent
%%%%%%%%%%%%%%%%%%%%%%%%%%%%%%%%%%%%%%%%%%%%%%%%%%%%%%%%%%%%%
% This is the abstract
%\begin{abstract}
%Ins
Recently an interesting multi-ferroic system PbVO$_{3}$ [Chem. Mater. 2004] 
was successfully prepared using a high-pressure and high-temperature 
technique. The crystallographic features were reported. In this note we 
concentrate on the theoretical XANES spectra by considering the K-edge 
of Vanadium. 
The tetragonality [c/a=1.229 at 300 K] of PbVO$_{3}$ is the largest
in the PbTiO$_{3}$ family of compounds. Thus one is led naturally
to examine the effect of the change of tetragonality and the axial
oxygen position on the electronic structure [i.e. XANES spectrum]. We study 
this effect in two ways. At a given temperature we vary the tetragonality 
and the axial oxygen position and quantify it in terms of XANES difference 
spectrum. Secondly, we compute the XANES spectra at three different 
temperatures, 90 K, 300 K, and 530 K and quantify the change in terms of 
the difference spectrum. We note that in this compound the tetragonality 
increases almost monotonically with temperature from 12 K to 570 K without 
transition to the cubic phase under ambient pressure. A key objective of the
current investigation is to gain an understanding of various absorption
features in the vicinity of K-edge of V, in terms of valence, local site 
symmetry, local coordination geometry, local bond distances, charge transfer,
and local projected density of states. We consider both the polarized and
the unpolarized XANES spectra. In short we have 
performed a local electronic study, which nicely complements the 
crystallographic features reported recently in PbVO$_3$.

% maybe should be in introduction
%The calculated XANES  spectra are remarkably self-consistent. We have 
%performed several cross-checks in addition to the Self-Consistent-Field [SCF] 
%cycles used in FEFF8 to obtain high level accuracy. It is to  be noted that 
%the calculations are based on SCF  one-electron Green's function approach, 
%where many-body effects are taken into account in terms of final-state 
%potentials and a complex energy-dependent self-energy. In this scattering 
%theoretic approach the structure in XANES is correlated with projected 
%density of states [pDOS]. 

%\end{abstract}
\end{minipage}
%\vskip 0.125cm 
%\pacs{78.20.-e, 78.30.-j, 74.76.Bz}
\end{center}
\vfill
\baselineskip=20pt
\normalsize
\newpage
\setcounter{page}{2}
%------------------------------------------------------------------------ 

%------------------------------------------------------------------------ 
% Section: 
%\section{Introduction}
% 19/1/2005	
\section{Introduction}
    Using high-pressure and high-temperature techniques, PbVO$_{3}$
has been synthesized by Belik et al.~\cite{ale04}, and 
Shpanchenko et al.~\cite{shp04}. 

    One of the most interesting feature of PbVO$_{3}$ is its 
large tetragonality [c/a=1.229 at 300 K] in comparison to that 
of PbTiO$_{3}$ [c/a=1.064 at 300 K]. The large
tetragonal distortion implies that it will have also a sizable
polarization, P$_{s}$. Indeed a rough calculation \cite{ale04}
employing the ionic model gives a value close to 100 $\mu$C/cm$^{2}$.
This is to be compared to P$_{s}$=81 $\mu$C/cm$^{2}$ at 300 K for 
PbTiO$_{3}$, P$_{s}$=26 $\mu$C/cm$^{2}$ at 300 K for BaTiO$_{3}$,
and P$_{s}$=37 $\mu$C/cm$^{2}$ at 543 K for KNbO$_{3}$, in the same
family of compounds. It is well-known and reasonable to expect that
{\em ferroelectricity} in ionic crystals is correlated with crystal
structure distortion. The basic reason is simple, polarization
is directly caused by the atomic displacement. A very well-known example
is PbTiO$_{3}$, a perovskite-type tetragonal crystal structure,
space group P4mm at 300 K and a simple cubic structure above
763 K. More interesting is the phenomenon, when a material exhibits
two or all three of the properties of (anti)ferroelectricity,
(anti)ferromagnetism, and (anti)ferroelasticity. These materials
are called multiferroics, a recent example is BiFeO$_{3}$ which
exhibits a PbTiO$_{3}$-type structure. Other examples are
YMnO$_{3}$, BiMnO$_{3}$, TbMnO$_{3}$, and  TbMn$_{2}$O$_{5}$.
It is clear that these materials contain a magnetic transition
metal ion [such as Mn, Fe] in conjunction with Bi$^{3+}$ and
Pb$^{2+}$.

It is well-known that as the region close to the x-ray near edge
is scanned in energy, the ejected photolectron sequentially 
probes the unoccupied electronic levels of materials. This
results in a fine x-ray absorption near edge structure [XANES],
within roughly 30-50 eV of the threshold. XANES thus contains
useful chemical and structural information. 

The main point is that this region is dominated by {\em strong}
photoelectron scattering. Thus it is highly non-trivial to
develop a code to take into account the multiple scattering
which dominate the XANES region.
Perhaps one of the most elegant XANES code is the FEFF8
series\cite{ank98,ank02}. This {\em ab initio} based on 
the {\em real-space multiple scattering}
[RSMS] and one of its several advantages is that it applies to both 
{\em periodic} and {\em aperiodic} systems.

Thus the purpose of this note is to concentrate on the
theoretical calculation using the updated version of FEFF8, i.e.
FEFF8.20 \cite{ank02}, for the PbVO$_{3}$, using the crystallographic
data reported recently by one of us \cite{ale04}. In particular we
calculate the near edge spectra of this system for several cases of 
interest. In addition we also study the near-edge
structure for the in-plane and out of plane cases.

Here we report on our results of PbVO$_{3}$, for the P4mm phase. 
The calculated XANES  spectra are remarkably self-consistent. We have 
performed several cross-checks in addition to the Self-Consistent-Field [SCF] 
cycles used in FEFF8 to obtain high level accuracy. It is to  be noted that 
the calculations are based on SCF  one-electron Green's function approach, 
where many-body effects are taken into account in terms of final-state 
potentials and a complex energy-dependent self-energy. In this scattering 
theoretic approach the structure in XANES is correlated with projected 
density of states [pDOS].

Another aim is to provide a concrete example of the application of XANES 
and DOS calculations using FEFF. Indeed, although FEFF is a powerful code, 
its self-consistency must be demonstrated by applications to real systems 
systematically. Moreover by actual calculations one can illustrate advantages
and disadvantages, and thus find ways of improving the code.

This paper is organized as follows, in the 
next section we outline some basic points about 
the multiple scattering calculations based on the FEFF8 code.
In section three the results and discussion of our study
of PbVO$_{3}$ are given using the FEFF8 code. The final
section contains the conclusions. 

%%%%%%%%%%%%%%%%%%%%%%%%%%%%%%%%%%%%%

\section{Multiple Scattering via FEFF8}

	It is well-known that XANES refers to roughly the 40-50 eV 
region near the edge of X-ray absorption spectroscopy [XAS].
The main point is that this region is dominated by {\em strong}
photoelectron scattering. Thus it is highly non-trivial to
develop a code to take into account the multiple scattering
which dominate the XANES region.
Perhaps one of the most elegant XANES code is the FEFF8
series\cite{ank98,ank02}. This {\em ab initio} based on 
the {\em real-space multiple scattering}
[RSMS] and one of its several advantages is that it applies to both 
{\em periodic} and {\em aperiodic} systems. In this approach the
self-consistent field [SCF] calculations for both the local electronic
structure and x-ray absorption spectra are implemented. It is important
to note that the full-multiple scattering [FMS] is taken into account
for a ``small cluster'' of atoms plus the higher-order multiple scattering 
from scatterers outside the said cluster. Some of its main advantages
are: SCF estimate of Fermi energy, orbital occupancy and charge transfer.

In this note we concentrate mainly on giving the $\mu(E)$ and the
difference $\mu(E)$ spectra. The value $\mu(E)$ is the main 
quantity in XANES.
Here we look at the basic definitions restricting ourselves
to brief comments. As is well-known
the primary quantities with which XANES calculations are concerned
are $\mu$,$\mu_{0}$, $\chi$, and $\rho_{li}(E)$\cite{ank98}
\begin{eqnarray}
\mu_{li}(E)=\mu_{li}^{0'}(E)[1+\chi_{li}^{'}(E)].
\label{f1}
\end{eqnarray}
%%%%%%%% 
We note that the prime is here for clarity since it
reminds us that it denotes final state quantities
in the {\em in the presence of screened hole}.
The central quantity in X-ray Absorption Spectroscopy [XAS] 
is the absorption coefficient $\mu(E)$. As is known there is a close and 
deep connection between XAS and {\em electronic structure}, which 
is indicated and implied by the resemblance of the contribution 
from a site, $i$ and
orbital angular momentum $l$ and the local $l$-projected
electronic density of states [LDOS] at site $i$
\begin{eqnarray}
%\rho_{li}(E)=\rho_{li}^{0'}(E)[1+\chi_{li}(E)].
\rho_{li}(E)=\rho_{li}^{0}(E)[1+\chi_{li}(E)].
\label{f2}
\end{eqnarray}
We provide, a concrete example of this inter-relationship between 
absorption and LDOS in Fig.~\ref{fig4}. It is clear that the main
features or peaks of the two quantities, absorption and LDOS
strongly resemble each other, as they should. Incidentally, this
provides an independent check on our calculation.

However, it is important to bear in mind that the
since core hole plays a significant role in calculation
of XAS, the similarity between XAS and LDOS cannot be 
regarded as an absolute.

	It is important to keep in mind that FEFF method starts
from the most fundamental quantity i.e. the Real Space Green's
Function and constructs the physical quantities of interest
from it \cite{ank98,ank02}. This is one of the code's main attraction 
since unlike band calculations it does not depend on symmetry. 
In this sense it is ideal for cluster physics\footnote{It is known 
that XANES and EXAFS signals are sensitive
to local structure. Indeed just as XRD is indicative of long-range
order, XANES and EXAFS carry signatures of short range order or
disorder.}. In the shorthand notation the MS expansion can be
written as \cite{ank98}
\begin{eqnarray}
G^{SC}=G^{0}tG^{0}+G^{0}tG^{0}tG^{0}+G^{0}tG^{0}tG^{0}tG^{0}+...,
\label{f3}
\end{eqnarray}   
where $G^{SC}$ represents the self-consistent Green's Function
and t the scattering t matrix. It is clear that in FMS we can implicitly
sum Eq.\ref{f3} to all orders by using matrix inversion, i.e.
\begin{eqnarray}
G^{SC}=G^{0}(1-tG^{0})^{-1},
\label{f4}
\end{eqnarray} 
as it follows simply from the form of  Eq.\ref{f3}.
%%%%%%%%%%%%%%%%%%%%%%%%%%%%%%%%%%%%%%%%%%%%%%%%%%%%%%
%\section{XANES Results}
%\section{}
%\subsection{Raw Data and Standard}
%\subsection{Main Edge}
%\subsection{Pre-edge Region}
%\subsection{Difference Spectra}

\section{Results and Discussion}
	Let us now give the results and analysis
of our calculated theoretical results using FEFF8.20.

The major input we use is the crystallographic and positional
information refinements recently given by one of us \cite{ale04}.
This information is available at 90 K, 300 K and 570 K and for
convenience is listed in Table I.

For the purposes of this note we consider the XANES results,
where we consider all the atoms upto the distance of 7 \AA, from 
the central or core atom, i.e. V. The exact enumeration is given
in Table II for the case, when the temperature is 90 K. The distances 
from V are calculated with Atoms [version 3.0beta10] 
program \cite{new01}. A total of 110 atoms including the central 
atom are included in the current XANES calculation,
Table II. The shell and atom numeration for temperatures 300 K
and 530 K are respectively given in Tables III and IV. From Tables II-IV,
several points are immediately clear. We can clearly see that V atom
is five-fold coordinated, with axial oxygen atoms asymmetrically 
situated. In other words, there is a very strong octahedral distortion
with one very short Vanadyl V-O distance. Thus the axial oxygen 
distances in PbVO$_{3}$ are much further apart compared to PbTiO$_{3}$.
This information is summarized as a comparison of PbVO$_{3}$ and 
PbTiO$_{3}$ in Table V. The oxidation state of the five-fold 
coordinated V can be readily be estimated by using the relation 
$Z=25.99-11.11 l_{V-OP}$\cite{sch00}
and the value of $V-OP=1.98583$ \AA from Table V, this gives
Z=3.9274. Thus V has an average valence of approximately 4+.
{\em We expect this strong five-fold coordination to effect
the electronic properties}, and this is shown to be the case
as we show below.

The main input file for FEFF8.2 is generated by Atoms
and analysis is done with Athena\cite{new01} [version 0.8.037].

Typical XANES spectra for the three different temperatures 90 K, 300 K and
530 K are given, respectively in Figs.~\ref{fig1}-\ref{fig3}.
These are labelled by $\mu_{u}$ for the unpolarized, by $\mu_{ab}$ 
for the polarized case representing the ab-plane and by $\mu_{c}$, when 
the E vector is parallel to the c-axis. 

In Fig.~\ref{fig4} the XANES spectrum is compared with the corresponding
projected density of states. The agreement is quite good, showing
the correctness of our calculations.

In Fig.~\ref{fig5} the XANES spectra for the two cases, I and II
are given using the 90 K input data. Intuitively, one expects
that if the axial oxygen position is changed, one would see the
consequence of this change in the local electronic structure, i.e.
the XANES spectra. In the example given in Fig.~\ref{fig5} we took
the apical oxygen fractional z-coordinate as $\delta z_{_{V-OA}}=0.1612$  
and labelled this as case I. For comparison we note that the corresponding
values of axial oxygen are $\delta z_{_{V-OA}}=0.1138$ at room temperature
for PbTiO$_{3}$ and $\delta z_{_{V-OA}}=0.2087$ and 
$\delta z_{_{V-OA}}=0.2102$, respectively at 90 K and 300 K for PbVO$_{3}$,
Table I. Thus, for illustration we took an intermediate value between
that of PbTiO$_{3}$ and  PbVO$_{3}$ for this quantity, to see how it 
would effect the electronic structure. Incidentally, one would also 
expect roughly this value for the fractional z-coordinate for the system 
PbV$_{0.5}$Ti$_{0.5}$O$_{3}$, which has not been fabricated yet. 
It would be useful to fabricate PbV$_{0.5}$Ti$_{0.5}$O$_{3}$, since it
could provide further insight between distortion and multi-ferroic
behaviour in these and related systems. 

We also expect that varying the tetragonality [c/a] at a given temperature
will also effect the XANES spectrum. Indeed this turns out to be the case.
This case is labelled as case II. Once again, for demonstration we took 
an intermediate between that of 
PbTiO$_{3}$ [a=b=3.905~\AA~,c=4.156~\AA~, c/a=1.06428 at 300 K]
and PbVO$_{3}$ [a=b=3.8033~\AA~,c=4.6499~\AA~, c/a=1.2226 at 90 K].
For case II we chose a=b=3.905~\AA~,c=4.4700~\AA~, c/a=1.1447 at 90 K.
 
The positional information [in energy units eV] of the features or 
peaks for Figs.~\ref{fig1}-\ref{fig3} and \ref{fig5} are given in 
Table~VI.

In order to quantify the change in XANES spectra, as a result of
the temperature change , varying the axial oxygen position and the
tetragonality at a given temperature, we calculate the difference 
spectra of the cases with respect to the spectra at 90 K. 

The results for difference spectra for the temperatures 300 K and 500 K 
data with respect to 90 K data are shown in Fig.~\ref{fig6}.

The difference spectra which result when the axial oxygen position
and tetragonality is varied are illustrated in Fig.~\ref{fig7}.

Let us consider first some general remarks as a guide to the reader on
Vanadium K-edge \cite{wong84} in the context of Vanadium oxides, in order 
to motivate the spectra shown in Figs.~\ref{fig1}-\ref{fig3} and \ref{fig5}.
It is known that V forms a series of oxides VO ,~V$_{2}$~O$_{3}$,~
V$_{4}$O$_{7}$,~V$_{2}$O$_{4}$ and V$_{2}$O$_{5}$ over a range
of formal oxidation states.
In V K-edge XANES/ELNES in these oxides, one observes a pre-edge
absorption feature which is strongest in V$_{2}$O$_{5}$, followed
by weak feature/shoulder on a rising absorption edge, which culminates
in a strong peak approximately 20 eV from the Fermi energy. It is
known that the strong peak is assigned to dipole-allowed transition
1s $\longrightarrow$ 4p, the weak shoulder/feature as a shakedown
transition representing the 1s $\longrightarrow$ 4p and the pre-edge
or near-edge feature near the threshold as a forbidden 
1s $\longrightarrow$ 3d. The features/peaks arising at higher energies
than the main 1s $\longrightarrow$ 4p transition are more difficult
to be specific about. These features may arise from transition to higher
np states, multiple scattering and/or shape resonances.
 
	From Figs.~\ref{fig1}-\ref{fig3} and \ref{fig5} we can clearly
see five features or peaks. The first peak P1 can be referred to as the
edge peak, since it just above the Fermi energy. This feature is present
in all the cases. However, there are some interesting points to note, 
regarding this peak. First of all, the dependence of this peak on the
distortion of the local structure around V atom, is evident from our
calculation of the different cases. It does not disappear with 
temperature, since the distortion is present at higher temperatures,
in contrast to what would happen in case of centrosymmetric structures.
The position of P1 also remains close to approximately 5469.1 eV, Table VI,
for all the cases considered here, with the exception of case II, where
it is slightly differently located at roughly 5469.4 eV. In order to
show and quantify that the distortion and ultimately multi-ferrocity
is caused by the asymmetric positions of the axial oxygens, we calculated
the polarized XANES data along the c-axis and ab-plane. It is clear
from Figs.~\ref{fig1}-\ref{fig3} that the main contribution to distortion
comes from the c-axis spectra. Thus the unpolarized data $\mu_{u}$, which 
is an average of all the polarizations, receives most of the contribution
for P1 from the c-axis. In contrast the ab-plane spectra exhibits P1
to a much lessor degree. These points are cleared further by examining P1
in the 530 K spectra, here P1 is quite weak [WF] for the ab-plane data, which
effects the polarization average data, so that after averaging P1 appears
as a weak [W] feature in the unpolarized spectra, in contrast P1 remains
a strong feature in the c-axis spectra.

The strong dependence of the XANES/ELNES on the local distortion is clear
not just from P1, but the entire spectrum. For example, P4 [the dominant
post edge peak at roughly 5487 eV, is nearly absent for the most 
non-centrosymmetric case, i.e. the c-axis spectra, Table VI and 
Figs.~\ref{fig1}-\ref{fig3}, where we
have labelled it as very weak feature [VWF]. P4 would be very large
for centrosymmetric case. In short, it anti-correlates with distortion,
in contrast to P1 which correlates with deviation from centrosymmetry.
On the basis of our data, we can also conclude that P5 also correlates
with distortion. The ''shakedown'' peaks P2 and P3 are quite robust
and only P2 slightly weakens at 530 K, Table VI.

Yet another interesting result of our calculation is that P1 arises,
due to the hybridization between the p and d orbitals of the absorber,
an effect which is forbidden for centrosymmetric structure but increased
by local distortions. This effect is known for PbTiO$_{3}$ \cite{ank98},
but for PbVO$_{3}$ it seems to be very pronounced. This is expected due
the much bigger tetragonality of the PbVO$_{3}$ compared to PbTiO$_{3}$.
But our calculations explicitly demonstrate it and further quantifies 
it in terms of the electronic near edge structure.

The results for the two cases I and II, are displayed in Fig.~\ref{fig5},
the five peaks/features are apparent again. The exact positions of the 
features are listed in Table VI. It is interesting to note that for case
I the peak P4 is dominant compared to P1, whereas it is reversed in case
II, Fig.~\ref{fig5}. Thus once again we see the anti-correlation of the
two peaks P1 and P4, as discussed above. Once again the ''shakedown'' 
peaks P2 and P3 are present. 

In order to further quantify the near edge electronic structure we have
taken the difference spectra, of all the unpolarized cases discussed here, 
relative to the 90 K spectra. The results are displayed in 
Figs.~\ref{fig6}-\ref{fig7}. The percentage change in tetragonality
in going from 90 K to 300 K and 530 K, Table I, is approximately
0.6\% and 0.9\% respectively, with respect to the former. This
translates into approximately 4\% for the 300 K case and about
10.5\% for the 530 K, as can be seen from the difference spectra,
Fig.~\ref{fig6}, by considering the maximum peak to peak variation.
This clearly shows that we can correlate small changes in tetragonality 
to the local electronic behaviour. Let us turn to Fig.~\ref{fig7},
where for case II, the tetragonality was chosen as c/a=1.1447, which
is 6.4 \% smaller than that of PbVO$_{3}$ relative to the latter, at
the same temperature of 90 K. The change in the corresponding difference
spectra is large, as expected, average variation being on the order of 
20\%. This shows that by taking tetragonality roughly between that
of PbTiO$_{3}$ and that of PbVO$_{3}$, we can account for the large
difference between their electric polarizations, which is incidentally
also on the order of 20 \%, albeit at different temperature of 300 K. 
For case I, where we attempted to simulate a behaviour between
PbTiO$_{3}$ and PbVO$_{3}$, by taking the z-coordinate of the axial
oxygen roughly between these two systems, keeping the tetragonality
at the same value as PbVO$_{3}$, we see an average percentage change
of 30 \% relative to the PbVO$_{3}$ at the same temperature. In particular
variation is large in region of peak P1, as expected in lieu of
earlier comments, since the more the deviation from perfect octahedral
symmetry, the bigger the effect, clearly the PbVO$_{3}$ at 90 K has
$\delta z_{_{V-OA}}=0.2087$ to be compared with $\delta z_{_{V-OA}}=0.1612$
for case I, at the same temperature. This represents a percentage
change of roughly 23 \% in the $\delta z_{_{V-OA}}$ value relative
to the measured PbVO$_{3}$ at 90 K. In order to see how this effects
the difference spectra, we can examine the P1 region in Fig.~\ref{fig7}
for case I. Incidentally the change is around 23 \%, which is consistent
with percentage change of $\delta z_{_{V-OA}}$.

The charge transfer and occupational orbital numbers are given
in Tables VII and VIII. These results can be used to analyze the
varying influence of the oxygen environment on the electronic
structure of different transition-metal atoms in general, and
in particular in the perovskite structure of our interest. Here
we only concentrate on a comparison between  PbVO$_{3}$ and 
PbTiO$_{3}$, leaving a more detailed comparison between several
related materials for future work. The case of PbTiO$_{3}$ has been 
calculated in some detail for the purposes of comparison, since
it possesses the largest tetragonality among the previous known
members of this family. We see that the charge transfer for
the oxygen atom in the  PbVO$_{3}$ changes by approximately by 28.9\% 
relative to the PbTiO$_{3}$ case at 300 K. For the Pb atom one see
a change of roughly 16.3 \% compared to that in the  PbTiO$_{3}$
system. We note that for each case, the net charge transfer
cancels to within $\pm 0.001$, which provides a useful and a simple
test for our calculations. 

The exact location of all the peaks is displayed in Table IX. Both
the positive and negative features are presented for obvious reasons.
We can immediately see that difference spectra shows the largest 
variations, and peaks roughly at the positions P1, P2, P3, P4 and P5, 
discussed above. This is not surprising, but clarifies the differential 
behaviour of each peak region for the unpolarized spectra given in   
Figs.~\ref{fig1}-\ref{fig3} and \ref{fig5}.

%%%%%%%%%%%%%%%%%%%%%Section %%%%%%%%%%%%%%%%%%%%%%%%%% 	     
\section{Conclusions}
We have for the first time given the XANES study of the
newly fabricated PbVO$_{3}$.
\begin{itemize}
\item{}The spectra for three important cases and the difference
spectra have been calculated and presented here. It is demonstrated
that electronic structure shows a calculable change in the temperature
range 90 - 530 K. This is expected since the c-lattice parameter 
increases almost linearly with temperature, whereas the a and b lattice
constants remain unchanged.
\item{} Roughly speaking all the spectra show five peaks or features in
the XANES region.  
\item{} We found that by taking polarization into account, we can
clearly see the origin and the dependence of peaks and features and
their sensitivities, on ab-plane or c-axis. This also allows us to trace
back the origin of peaks/features in the unpolarized spectra to the
ab-plane and/or c-axis.  
\item{} Charge transfers and electronic orbital occupation numbers
have been calculated for several cases of interest. We have also 
given a comparison of charge transfer and orbital occupation numbers 
between PbVO$_{3}$ and  PbTiO$_{3}$. For example, we find a change of
approximately 28.9\% in the charge transfer of the oxygen atom in going
from the PbVO$_{3}$ to the PbTiO$_{3}$ with respected to the latter,
for the 300 K data.
\item{} The difference spectra confirms the previous observations and
results, and further correlates the percentage changes in tetragonality
and $\delta z_{_{V-OA}}$ directly to the variations observed in the
difference spectra. We can also account roughly for the change in 
electric polarization between PbVO$_{3}$ and  PbTiO$_{3}$, by looking
at how changes in tetragonality and axial-oxygen deviation from ideal 
position correlate to the variation in the near edge electronic structure.
The difference spectra allows us approximately to translate these into
percentage changes.
\end{itemize}
From these calculation and analysis
we can conclude that is possible to quantify the  
electronic structure of PbVO$_{3}$ by using the
near edge structure.

%\end{itemize} 
%%%%%%%%%%%%%%%%%%%%%%%%%%%%%%%%%%%%%%%%%%%%%%%%%%%%%%%%%%%%
%\begin{enumerate}
%\item{}
%\end{enumerate} 
%%%%%%%%%%%%%%%%%%%%%%%%%%%%%%%%%%%%%%%%%%%%%%%%
%In a metallic system close to a metal-insulator transition
%the electronic properties play an important and a crucial role.
%%%%%%%%%%%%%%%%%%%%%%%%%%%%%%%%%%%%%%%%%%%
%\newpage
\section*{Acknowledgments}
The Sher Alam's work is supported by Y.Matsui's Crystal 
Structure and Analysis Group at the Advanced Material 
Laboratory [NIMS] and MONBUSHO via the JSPS invitation program.

%%%%%%%%%%%%%%%%%%%%
%Tables
\newpage
\hspace{-.5in}Table I: Crystallographic Data at 90 K, 300 K and
530 K for PbVO$_{3}$.\\
\\
%%\begin{table}
\begin{tabular} {|l|c|c|c|c|c|c|c|r|}\hline
%\multicolumn{4}{c}
Temperature & Atom & a \AA & c \AA & $\frac{c}{a}$ & x & y & z\\
\hline
\hline
90 K & Pb & 3.80329(6) & 4.64989(10)& 1.2226 &0   &0    &0\\ 
     & V  &            &            & &1/2  &1/2  &0.5677(3)\\
     & O1 &            &            & &1/2  &1/2  &0.2087(15)\\
     & O2 &            &            & &1/2  &0    &0.6919(9)\\
\hline
300 K & Pb & 3.80391(5)& 4.67680(8) &1.2295 &0   &0    &0\\ 
     & V  &            &            &  &1/2  &1/2  &0.5668(4)\\
     & O1 &            &            &  &1/2  &1/2  &0.2102(16)\\
     & O2 &            &            &  &1/2  &0    &0.6889(10)\\
\hline
530 K & Pb & 3.80721(5) & 4.69819(9) &1.2340& 0   &0    &0\\ 
     & V  &            &            & &1/2  &1/2  &0.5674(4)\\
     & O1 &            &            & &1/2  &1/2  &0.2102(15)\\
     & O2 &            &            & &1/2  &0    &0.6915(10)\\
\hline
\hline
%\label{table1}
\end{tabular}
%\end{table}
\\
\\
\\
%\newpage
\hspace{-.5in}Table II: Shell and atom enumeration in the P4mm,
with V as a core atom at 90 K.\\
\\
%%\begin{table}
\begin{tabular} {|l|c|c|r|}\hline
%\multicolumn{4}{c}
Shell & Atom & No.of Atoms & distance from V \\
\hline
\hline
1 & O1 & 1 & 1.66931 \\
2 & O2 & 4 & 1.98741 \\
3 & O1 & 1 & 2.98058 \\
4 & Pb & 4,4 & 3.35757,3.76839 \\
5 & V  & 4 & 3.80329 \\
6 & O1 & 4 & 4.15351 \\
7 & O2 & 8,4 & 4.29125,4.49449 \\
8 & V    & 2   & 4.64989 \\
9 & O1   & 4   & 4.83207 \\
10& V    & 4   & 5.37866\\
11 & O2    & 4 & 5.56256 \\
12 & O1    & 4 & 5.63175 \\
13 & O2    & 4,8 & 5.73409,5.88774 \\
14 & V  & 8 & 6.00720 \\
15 & O1  & 4,1 & 6.14930,6.31920 \\
16 & Pb   & 8,8     & 6.34060,6.56740 \\
17 & O2   & 8,8   & 6.73848,6.88075\\
\hline
\hline
%\label{table2}
\end{tabular}
%\end{table}
\\
\\
\\
%\\
\newpage
\hspace{-.5in}Table III: Shell and atom enumeration in the P4mm,
with V as a core atom at 300 K.\\
\\
%%\begin{table}
\begin{tabular} {|l|c|c|r|}\hline
%\multicolumn{4}{c}
Shell & Atom & No.of Atoms & distance from V \\
\hline
\hline
1 & O1 & 1 & 1.66775 \\
2 & O2 & 4 & 1.98583 \\
3 & O1 & 1 & 3.00905 \\
4 & Pb & 4,4 & 3.36742,3.77647 \\
5 & V  & 4 & 3.80391 \\
6 & O1 & 4 & 4.15345 \\
7 & O2 & 8,4 & 4.29107,4.52490 \\
8 & V    & 2   & 4.67680 \\
9 & O1   & 4   & 4.85017 \\
10& V    & 4   & 5.37954\\
11 & O2    & 4 & 5.58187 \\
12 & O1    & 4 & 5.63213 \\
13 & O2    & 4,8 & 5.73437,5.91138 \\
14 & V  & 8 & 6.02845 \\
15 & O1  & 4,1 & 6.16391,6.34455 \\
16 & Pb   & 8,8     & 6.34658,6.57276 \\
17 & O2   & 8,8   & 6.75478,6.88133\\
\hline
\hline
%\label{table3}
\end{tabular}
%\end{table}
\\
\newpage
\hspace{-.5in}Table IV: Shell and atom enumeration in the P4mm,
with V as a core atom at 530 K.\\
\\
%%\begin{table}
\begin{tabular} {|l|c|c|r|}\hline
%\multicolumn{4}{c}
Shell & Atom & No.of Atoms & distance from V \\
\hline
\hline
1 & O1 & 1 & 1.67819 \\
2 & O2 & 4 & 1.99089 \\
3 & O1 & 1 & 3.02000 \\
4 & Pb & 4,4 & 3.37316,3.78862 \\
5 & V  & 4 & 3.80721 \\
6 & O1 & 4 & 4.16067 \\
7 & O2 & 8,4 & 4.29633,4.53410 \\
8 & V    & 2   & 4.69819 \\
9 & O1   & 4   & 4.85955 \\
10& V    & 4   & 5.38421\\
11 & O2    & 4 & 5.61384 \\
12 & O1    & 4 & 5.63968 \\
13 & O2    & 4,8 & 5.74051,5.92055 \\
14 & V  & 8 & 6.04713 \\
15 & O1  & 4 & 6.17334 \\
16 & Pb  & 8 & 6.35358\\
17 & O1  & 1 & 6.37638\\
18 & Pb   & 8  & 6.58357 \\
19 & O2   & 8,8   & 6.78307,6.88827\\
\hline
\hline
%\label{table4}
\end{tabular}
%\end{table}
\\
\\
\\
\hspace{-.5in}Table V: Typical oxygen distances in PbVO$_{3}$ versus
PbTiO$_{3}$ at 300 K, from Pb and M [V,Ti] atoms. The oxygen distances 
for the two cases I and II at 90 K considered here are also given, in
addition to the cases of PbVO$_{3}$ for 90 K and 530 K.
All distances are in \AA~~units.\\
\\
%%\begin{table}
\begin{tabular} {|l|c|c|c||c|c|c|r|}\hline
%\multicolumn{5}{c}
Material &Pb-OA & Pb-OP &M-OA[short] \AA & M-OP \AA & M-OA[long] \AA & Z \\
\hline
\hline
PbVO$_{3}$  &2.86378 & 2.38614 & 1.66775 & 1.98583 & 3.00905 & 3.9274\\
PbTiO$_{3}$ &2.80146 & 2.51937 & 1.76713 & 1.97916 & 2.38887 & 4.0015\\
Case I      &2.79184 &2.38090&1.89018 & 1.98741 & 2.75972  & 3.9099 \\ 
Case II     &2.91458 &2.38934&1.60473 & 2.02989 & 2.86527 & 3.4379\\        
PbVO$_{3}$ [90 K]&2.85907 &2.38090&1.66931 & 1.98741 & 2.98059  & 3.9099 \\
PbVO$_{3}$ [530 K]&2.86752 &2.39258&1.67820 & 1.99089 & 3.0200  & 3.8712 \\
\hline
\hline
%\label{table5}
\end{tabular}
%\end{table}
\\
\\
\newpage
\hspace{-.5in}Table VI: Features/Peaks for Figs.~\ref{fig1}-\ref{fig3}
and Fig.~\ref{fig5}. All the positions are given in eV.
\\
%%\begin{table}
\begin{tabular} {|l|c|c|c|c|r|}\hline
%\multicolumn{6}{c}
Case   & P1   &  P2   &  P3     & P4  & P5\\
\hline
90 K,$\mu_{u}$ & 5469.087 & 5475.655 &5480.532 
& 5487.475 & 5493.943\\
             &          & 5477.341 &       
&          &         \\
90 K,$\mu_{ab}$ & 5469.087 & 5475.655 &5480.532 
& 5487.475 & VWF \\
90 K,$\mu_{c}$ & 5469.087 & 5475.655 &5480.532 
& VWF & 5493.943 \\
% VWF(5488.436)
\hline
\hline
300 K,$\mu_{u}$ & 5469.084 & 5475.652 &5480.529 
& 5487.472 & VWF\\
             &          & 5477.338 &       
&          &         \\
300 K,$\mu_{ab}$ & 5469.084 & 5475.652 &5480.529 
& 5487.472 & VWF \\
300 K,$\mu_{c}$ & 5469.084 & 5476.757 &5480.529 
&  VWF & 5493.94 \\
%VWF(5488.383)
\hline
\hline
530 K,$\mu_{u}$ & 5469.076 &VW(5476.749) &5481.216 
& 5488.010 & VWF\\
530 K,$\mu_{ab}$ & 5469.076 & W(5476.442) &5481.216 
& 5487.464 & VWF \\
530 K,$\mu_{c}$ & 5469.076 & 5476.749 & 5480.521
&VWF & 5492.951 \\
%VWF(5487.477)
\hline
\hline
case I & 5469.035 & 5479.146 &5481.175 & 5487.424 & 5496.948\\
\hline
\hline
case II& 5469.387 & 5475.131 &5479.855 & 5486.627 & 5495.962\\
\hline
\hline
%\label{table6}
\end{tabular}
%\end{table}
\\
\\
%\newpage
\hspace{-.5in}Table VII: Charge Transfer and Occupation orbital
numbers for the polycrystalline PbVO$_{3}$ at 90 K, 300 K and 530 K.
For comparison the case of polycrystalline PbTiO$_{3}$ at 300 K is given.
\\
\\
%%\begin{table}
\begin{tabular} {|l|c|c|c|c|r|}\hline
%\multicolumn{6}{c}
 Atom Type   & l character & 90 K      & 300 K     & 530 K  
& PbTiO$_{3}$ [300 K]\\
\hline
   O         & s   & 1.879   & 1.879  & 1.881 &~1.882,~~~~~1.888\\
             & p   & 4.357   & 4.355  & 4.361 &~4.448,~~~~~4.423\\
             & d   & 0.000   & 0.000  & 0.000 &~0.000,~~~~~0.000\\
Charge Transfer&   & -0.236  & -0.234 & -0.241&-0.329,~~~~-0.312 \\
Atoms in cluster&  & 57      &   57     &  57 &~~~53,~~~~~~~~15\\
\hline
V, Ti        & s   & 0.452   & 0.451  & 0.453 &~0.429,~~~~~0.439\\
             & p   & 6.653   & 6.653  & 6.655 &~6.663,~~~~~6.706\\
             & d   & 3.798   & 3.795  & 3.787 &~2.643,~~~~~3.792\\
Charge Transfer&   & 0.099   & 0.098  & 0.106 &~0.264,~~~~~0.081\\
Atoms in cluster&  &   49    & 45     &  45   &~~~37,~~~~~~~~15\\
\hline
V$_{core}$, Ti$_{core}$ & s  & 0.454 & 0.454  & 0.455 &~0.435,~~~~~0.439\\
             & p  & 6.674   & 6.673  & 6.675 &~6.690,~~~~~6.706\\
             & d  & 4.902   & 4.922  & 4.894 &~3.778,~~~~~3.792 \\
Charge Transfer&  & -0.037  & -0.037 & -0.031&~0.095,~~~~~0.081\\
Atoms in cluster& & 65      &   65   &  57 &~~~45,~~~~~~~~15\\
\hline
\hline
Pb           & s  & 1.756   & 1.765  & 1.763 &~~~1.835,~~~1.834\\
             & p  & 1.300   & 1.298  & 1.285 &~~~1.141,~~~1.136\\
             & d  & 10.334  & 10.333  & 10.322 &~~~10.301,~~~10.331\\
Charge Transfer&  & +0.610  & +0.605 & 0.617 &~~~0.723,~~~0.699\\ 
Atoms in cluster& &  46     &    46    &  46 &~~~42,~~~~~~~~21\\
\hline
\hline
\end{tabular}
%\end{table}
\\
%\newpage
\hspace{-.5in}Table VIII: Charge Transfer and Occupation orbital
numbers for the polycrystalline PbVO$_{3}$ at for case I and II.
For comparison the case of polycrystalline PbTiO$_{3}$ at 300 K is given.
\\
\\
%%\begin{table}
\begin{tabular} {|l|c|c|c|r|}\hline
%\multicolumn{6}{c}
 Atom Type   & l character & case I at 90 K  & case II at 90 K 
& PbTiO$_{3}$ [300 K]\\
\hline
   O         & s   & 1.894   & 1.870   &~1.882,~~~~~1.888\\
             & p   & 4.442   & 4.315  &~4.448,~~~~~4.423\\
             & d   & 0.000   & 0.000  &~0.000,~~~~~0.000\\
Charge Transfer&   & -0.335  & -0.185 &-0.329,~~~~-0.312 \\
Atoms in cluster&  & 57      &   57     &~~~53,~~~~~~~~15\\
\hline
V, Ti        & s   & 0.451   & 0.439  &~0.429,~~~~~0.439\\
             & p   & 6.685   & 6.638  &~6.663,~~~~~6.706\\
             & d   & 3.612   & 3.869  &~2.643,~~~~~3.792\\
Charge Transfer&   & 0.249   & 0.052  &~0.264,~~~~~0.081\\
Atoms in cluster&  &   49    & 45     &~~~37,~~~~~~~~15\\
\hline
V$_{core}$, Ti$_{core}$ & s  & 0.452 & 0.443 &~0.435,~~~~~0.439\\
                        & p  & 6.693   & 6.662  &~6.690,~~~~~6.706\\
                        & d  & 4.779   & 4.971  &~3.778,~~~~~3.792 \\
Charge Transfer         &    & 0.088  & -0.069 &~0.095,~~~~~0.081\\
Atoms in cluster        &    & 65     &   65   &~~~45,~~~~~~~~15\\
\hline
\hline
Pb           & s  & 1.767   & 1.752  &~~~1.835,~~~1.834\\
             & p  & 1.145   & 1.408  &~~~1.141,~~~1.136\\
             & d  & 10.334  & 10.339 &~~~10.301,~~~10.331\\
Charge Transfer&  & +0.756  & +0.502 &~~~0.723,~~~0.699\\ 
Atoms in cluster& &  46     &    43  &~~~42,~~~~~~~~21\\
\hline
\hline
\end{tabular}
%\end{table}
\\
\newpage
\hspace{-.5in}Table IX: Peak positions in eV, of the difference
spectra displayed in Figs.~\ref{fig6}-\ref{fig7}.
\\
\\
%%\begin{table}
\begin{tabular} {|l|c|c|c|r|}\hline
%\multicolumn{6}{c}
 Peak Type   & $\delta \mu_{90-300}$ & $\delta \mu_{90-530}$ 
& $\delta \mu_{90-case1}$ & $\delta \mu_{90-case2}$\\
\hline
   +ve       &  & 5468.787  &  & 5468.787 \\
\hline
   +ve       &  &   & 5469.087 &  \\
\hline
   +ve       & 5469.687 &   &  &  \\
\hline
   +ve       & 5470.626 &   &  &  \\
\hline
   +ve       &  & 5475.131  & 5475.131 &  \\
\hline
  +ve       & 5477.341 & 5477.341  &  & 5477.341 \\
\hline
  +ve       &  &   & 5480.532 &  \\
\hline
  +ve       & 5481.227 &   &  & 5481.227 \\
\hline
 +ve       &  &   & 5484.989 &  \\
\hline
+ve       &  & 5486.627  &  &  \\
\hline
+ve       & 5488.342 &   &  &  \\
\hline
+ve       &  &   & 5492.962 &  \\
\hline
 -ve       & 5468.687 & 5468.187  &  &  \\
\hline
 -ve       &  & 5469.387  &  & 5469.687 \\
\hline
 -ve       & 5474.626 &   &  & 5474.626 \\
\hline
 -ve       &  & 5476.198  &  &  \\
\hline
 -ve       &  &   & 5478.56 &  \\
\hline
 -ve       &  &   &  & 5479.198 \\
\hline
 -ve       &5480.532  &   &  &  \\
\hline
 -ve       &  & 5481.941  & 5481.941 &  \\
\hline
 -ve       &  &   &  & 5484.199 \\
\hline
 -ve       &  &   &5488.342  & \\
\hline
 -ve       &  & 5491.057  &  &  \\
\hline
 -ve       & 5492 &   &  &  \\
\hline
 -ve       &  &   &  & 5497 \\
\hline
\hline
\end{tabular}
%\end{table}
\\

%%%%%%%%%%%%%%%%%%
\newpage
%%%%%%%%%%%%%%%%%%%Ins%%%%%%%%%%%%%Figures%%%%%%%%%%%%%%
%\section*{Figure Captions}
%\begin{center}
%Figure Captions
%\end{center}
\begin{figure}
\caption{XANES $\mu$ data generated with FEFF8.2 for the
$p4mm$ phase of PbVO$_{3}$ at the V K-edge. Shown are the 
unpolarized, polarized with E$||$ab and E$||$c data, when the
temperature is 90 K.}
\label{fig1}
\end{figure}
%%%%%%%%%%%%%%%%%%%%%%%%%%%%%%%%%%%%%%%%%%%%%%%%%%%%%%%%%%%
\begin{figure}
\caption{Same as Fig.~\ref{fig1}, for the 300 K case.}
\label{fig2}
\end{figure}
%%%%%%%%%%%%%%%%%%%%%%%%%%%%%%%%%%%%%%%%%%%%%%%%%%%%%%%%%%
\begin{figure}
\caption{Same as Fig.~\ref{fig1}, for the 530 K case.}
\label{fig3}
\end{figure}
%%%%%%%%%%%%%%%%%%%%%%%%%%%%%%%%%%%%%%%%%%%%%%%%%%%%%%%%%%
\begin{figure}
\caption{XANES $\mu$ 90 K data generated with FEFF8.2 for the
$p4mm$ phase of PbVO$_{3}$  at the V K-edge, and compared to the 
corresponding projected density of states.}
\label{fig4}
\end{figure}
%%%%%%%%%%%%%%%%%%%%%%%%%%%%%%%%%%%%%%%%%%%%%%%%%%%%%%%%
\begin{figure}
\caption{XANES $\mu$ data generated with FEFF8.2 for the
$p4mm$ phase of PbVO$_{3}$ at the V K-edge at 90 K for the two cases
I and II.}
\label{fig5}
\end{figure}
%%%%%%%%%%%%%%%%%%%%%%%%%%%%%%%%%%%%%%%%%%%%%%%%%%%%%%%%%%%%

\begin{figure}
\caption{Difference $\mu$ [$\mu_{_{90}}-\mu_{_{i}}$ i=300,530] spectra
of temperatures 300 K and 530 K with respect to the spectra
at temperature 90 K.}
\label{fig6}
\end{figure}
%%%%%%%%%%%%%%%%%%%%%%%%%%%%%%%%%%%%%%%%%%%%%%%%%%%%%%%%%%%%%
\begin{figure}
\caption{Difference $\mu$ [$\mu_{_{90}}-\mu_{_{case i}}$ i=I,II] spectra
for the two cases I and II at 90 K with respect to the spectra at 
temperature 90 K.}
\label{fig7}
\end{figure}
%%%%%%%%%%%%%%%%%%%%%%%%%%%%%%%%%%%%%%%%%%%%%%%%%%%%%%%%%%%%
\end{document}